\documentclass[aps,prd,preprintnumbers,groupedaddress,nofootinbib,amssymb,notitlepage,eqsecnum]{revtex4-1}
\usepackage{here}
\usepackage{graphicx}
\usepackage{amsmath}
\usepackage{bm}
\usepackage{color}
\usepackage[dvipsnames]{xcolor}
\usepackage[utf8]{inputenc}
\usepackage{amsfonts}
\usepackage{placeins}
\definecolor{refs}{RGB}{245,156,74}
\usepackage[colorlinks=true,hyperfootnotes=true,citecolor=cyan]{hyperref}
\usepackage{comment}
\usepackage{amsmath}
\usepackage{tikz}
\usetikzlibrary{decorations.pathreplacing}
\usepackage{listings}
\definecolor{codebg}{RGB}{245,245,245}
\definecolor{codeblue}{RGB}{30,100,180}

\usepackage{enumitem}

\newcommand{\thetadm}{\theta_{\rm dm}}
\newcommand{\deltadm}{\delta_{\rm dm}}
\newcommand{\thetab}{\theta_{\rm b}}
\newcommand{\deltab}{\delta_{\rm b}}

\newcommand{\dd}{{\rm d}}

\newcommand{\mH}{{\mathcal H}}

\newcommand{\be}{\begin{equation}}
\newcommand{\ee}{\end{equation}}
\newcommand{\bea}{\begin{eqnarray}}
\newcommand{\eea}{\end{eqnarray}}

\allowdisplaybreaks

\begin{document}

\title{Dark matter-baryons elastic coupling}

\author{
David Figueruelo$^{1,2}$, Florencia Anabella Teppa Pannia$^{3,4,2}$}

\affiliation{
$^{1}$ {\it Department of Theoretical Physics, University of the Basque Country UPV/EHU, 48080 Bilbao, Spain.}\\
$^{2}${\it Institute of Theoretical Astrophysics, University of Oslo, N-0315 Oslo, Norway.}\\
$^{3}$ {\it Departamento de Matem{\'a}tica Aplicada a la Ingenier{\'i}a Industrial, Universidad  Polit{\'e}cnica de Madrid, E-28006 Madrid, Spain.} \\
$^{4}$ {\it Instituto Universitario de F\'isica Fundamental y Matem\'aticas~(IUFFyM), Universidad de Salamanca, E-37008 Salamanca, Spain.} 
}

\begin{abstract}

We consider a cosmological model where dark matter and baryons interact via a pure momentum transfer. We present the foundations of the interaction and derive its main equations. In this framework, only the Euler equations for the two interacting components differ from those in the standard scenario.
We also explain how this interaction relates with the common Tight-Coupling Approximation schemes typically used in many Boltzmann solver. After that, we implement the model in CLASS and CAMB Boltzmann solvers, which relay on those Tight-Coupling Approximation schemes and in SymBoltz, which is free from any approximation scheme. We demonstrate full consistency among all the implementations in the difference codes. We demonstrate how this type of interactions are useful regarding the $\sigma_8$ tension by producing a reduced clustering on smaller scales. Using cosmic microwave background, baryon acoustic oscillations, supernovae type Ia  data and low-redshift constraints from  Sunyaev-Zeldovich  cluster counts and from  DES year 6 results, we show the interacting scenario is mildly preferred when those low-redshift datasets are included.

\end{abstract}

\date{\today}

\maketitle

\section{Introduction}

Over the last decades, cosmology has evolved into a mature discipline, entering what is now commonly referred to as the era of \textit{precision cosmology}. This achievement has been made possible not only by the vast amount of observational data collected but, above all, by the unprecedented precision of modern datasets. Current cosmological probes span a wide range of observables tracing different epochs and scales throughout the history of the Universe. We have data from the comic microwave background~\cite{Planck:2018vyg}, baryons acoustic oscillations~\cite{DESI:2025zgx}, Lyman-$\alpha$~\cite{eBOSS:2019qwo,eBOSS:2019ytm}, cosmic chronometers~\cite{Moresco:2024wmr},  supernovae Ia~\cite{Scolnic:2021amr,Rubin:2023jdq,DES:2024jxu} or large scale structure~\cite{Giblin:2020quj,DES:2026fyc}, among many others. This plethora of datasets with their enormous precision has allowed us to build the so-called standard model of cosmology, the $\Lambda$ Cold Dark Matter model ($\Lambda$CDM). 

However, despite its remarkable success, $\Lambda$CDM is still far from representing the final word in cosmology. Precision does not necessarily imply completeness, the best probe of it being the lack of a fundamental description for the dark sector of the Universe. Its first and currently most abundant component, the cosmological constant $\Lambda$, is typically  assumed to be a constant contribution to the energy content of the Universe, arising from the standard action of General Relativity. However, its physical nature remains far from being fully understood, and it continues to pose significant theoretical challenges \cite{Martin:2012bt}. Even its mere fact of being a constant has been potentially challenged by recent results in baryons acoustic oscillations~\cite{DESI:2024mwx, DESI:2025zgx}. Regarding the second component, dark matter, its situation is not substantially different. We know it should behave as matter with an interaction via gravity, while with respect to the other three fundamental interactions it should be very, if any, weakly interacting. Apart from that, we know it should be non-relativistic. Even if the previous situation is not ideal, an even more profound puzzle hovers over the field in the form of the data tensions~\cite{CosmoVerseNetwork:2025alb}. The most statistically significant is the Hubble tension, which manifests as a $\sim5\sigma$ discrepancy in the inferred present-day expansion rate of the Universe when determined from early- or late-time cosmological probes. A second, less pressuring tension is the $\sigma_8$ (or $S_8$) tension \cite{Pantos2026,CarlosGG2024}. The parameter $\sigma_8$ quantifies the clustering on spheres of radii $8 h^{-1}\; \text{Mpc}$, while $S_8$  is used to avoid the known degeneracy between $\sigma_8$ and $\Omega_{\rm m}$, retaining its identity of a measure of clustering. When inferred from early-universe probes, which require integrating the entire cosmic evolution up to the present time within a given cosmological model, a larger value is obtained than that inferred from local probes measuring the clustering of the universe around us. 

A first answer to the previous tensions may emerge from improving the calibrations, yet to be detected systematics, or improved techniques with future experiments. 
Indeed, recent weak-lensing analyses with improved photometric redshift calibration and shear modelling have substantially reduced the reported $S_8$ \cite{Pantos2026}. 
Nevertheless, the absence of a fundamental description of the dark sector continues to motivate the exploration of new physics. 
One of the possible paths is having interactions, in particular interactions that involve one or both of the dark components (see Ref.~\cite{Wang:2024vmw} for a review). A particular class of interactions are the so-called pure momentum transfer or elastic interactions. In such scenarios, a fluid with pressure couples a pressureless matter fluid which acquires an effective pressure that counteracts the gravitational collapse. Consequently, clustering is reduced for that pressureless matter component, and hence the value of $\sigma_8$ is reduced, reconciling early and late time probes. One of the first appearances of these models dates back to Ref.~\cite{Simpson:2010vh}, but after that several models have been suggested involving different components and from different approaches~\cite{Pourtsidou:2013nha,Boehmer:2015sha,Pourtsidou:2016ico,Linton:2017ged,Asghari:2019qld,Chamings:2019kcl,Amendola:2020ldb,BeltranJimenez:2020iyx,BeltranJimenez:2020qdu,Figueruelo:2021elm,Linton:2021cgd,BeltranJimenez:2021wbq,Liu:2023mwx,Cruickshank:2025chm,Aoki:2025bmj,Jensko:2026taf,BeltranJimenez:2026ymd}. Even the non-linear formation of structures has been studied for some of them~\cite{Baldi:2014ica,Baldi:2016zom,Ferlito:2022mok,BeltranJimenez:2025yad}. The defining property of pure momentum transfer interactions is that they leave the background dynamics unchanged while reducing the values of the parameters $\sigma_8$ and $S_8$. Consequently, they can alleviate the $\sigma_8$ tension without worsening the $H_0$ tension and, when supplemented by an appropriate mechanism to address the latter, may offer a consistent way to relieve both tensions simultaneously.

In previous approaches, some of the authors of this work have proposed scenarios in which dark energy couples either to dark matter, the model called $\alpha$CDM~\cite{Figueruelo:2021elm}, or to baryons, the model called $\beta$CDM~\cite{BeltranJimenez:2020iyx}, through an elastic interaction formally identical to Thomson scattering at first order in perturbation theory. More generally, this class of interactions has been referred to as Covariantised dark Thomson-like scattering \cite{BeltranJimenez:2021wbq}. In these scenarios, the interaction becomes relevant when the pressure component, dark energy, contributes significantly to the cosmic budget, that is, in the late Universe. In that regime, the pressure of dark energy induces an effective pressure in dark matter ($\alpha$CDM) or in baryons ($\beta$CDM) via the momentum transfer, thus reducing its clustering and addressing the tension $\sigma_8$ / $S_8$. In this work, we adopt the same philosophy and, drawing again inspiration from Thomson scattering, we explore a similar interaction now between dark matter and baryons. In this case, neither component possesses intrinsic pressure. However, in the early universe baryons are tightly coupled to photons, which do exert an effective pressure through Thomson scattering. We therefore consider an interaction formally analogous to Thomson scattering, which is relevant at early times due to this effective baryonic pressure.  
Our model is then closely related to the dark matter–proton interactions studied in Refs.~\cite{Gluscevic2018,Boddy2018}, where momentum exchange can transmit part of the pressure support of the photon--baryon plasma to dark matter, suppressing the growth of small-scale perturbations. While those works describe the interaction through a microscopic dark matter--baryon scattering cross section, our model parameter $\gamma$ provides an effective phenomenological description of the momentum transfer.

We assume the interaction to be elastic, at least up to the first order in cosmological perturbations theory, ensuring that the background cosmology remains identical to that of the fiducial $\Lambda$CDM 
model.\footnote{Although the fiducial cosmology adopted in the aforementioned $\alpha$CDM and $\beta$CDM models was $w$CDM, in this work we consider the simplest $\Lambda$CDM scenario. In contrast to the previous cases, the dark matter--baryon interaction does not require a dynamical dark energy component. The extension to the more general $w$CDM framework is straightforward; however, such an analysis is beyond the scope of the present work.
} 
The first question we address is whether the effective pressure acquired by baryons through Thomson scattering can sufficiently suppress the clustering of the coupled matter component via pure momentum exchange. Subsequently, we investigate whether the current observational data favour the presence of such interaction and assess its potential impact on the $\sigma_8$/$S_8$ tension.

Before proceeding, we emphasize that the interaction considered in this work is introduced as an effective phenomenological description of a possible coupling between dark matter and baryons in the cosmological regime. 
We do not assume that this parametrization arises from a complete microscopic theory, nor that it can be straightforwardly extrapolated to local or high-energy environments. Accordingly, our analysis is deliberately restricted to the cosmological evolution of the Universe, where such an effective interaction may capture the relevant large-scale dynamics. A viable fundamental realization could involve additional mechanisms that suppress or modify the interaction outside the cosmological regime, thereby remaining consistent with existing constraints from laboratory experiments, collider searches, and local gravity tests. The construction of such a microscopic framework, however, lies beyond the scope of the present work.

This paper is organised as follows. In Section~\ref{sec:model} we present the model and its relevant equations. In Section~\ref{sec:numerical} we present the numerical implementation of the model and its effects in different observables. After that, in Section~\ref{sec:constraints} we display the results from the Markov Chain Monte Carlo analysis using several datasets. Finally in Section~\ref{sec:conclusions} we present the conclusions of the work.

\section{Momentum transfer interactions: Dark Matter - Baryons case}
\label{sec:model}

We assume that the Universe can be described by a background metric  of the flat Friedman-Lema\^itre-Roberton-Walker type, whose line element is  
\be
\dd s^2=a^2(\tau)\Big(-\dd \tau^2+\dd \vec{x}^2\Big)\,,
\ee 
where $\tau$ is the conformal time, $\vec{x}$ are the comoving coordinates and $a(t)$ is the scale factor.
The components of the Universe can be described in terms of perfect fluids with the total energy-momentum tensor given by 
\begin{equation}
T^{\mu\nu}=\sum_i\Big[(\rho_{i} +p_{i})u_i^\mu u_i^{\nu} +g^{\mu\nu}\, p_{i}\Big]\;,
\end{equation}
where $\rho_{i}$, $p_{i}$ and $u_i^\mu$ are the energy-density, pressure, and the 4-velocity of the $i$-th component and the sum runs over all the components of the Universe.   
We consider a dark energy component described by a cosmological constant, $\Lambda$, and assume that all components are covariantly conserved, i.e., $\nabla_\mu T_i^{\mu\nu}=0$, except for dark matter and baryons, which exchange momentum through a pure momentum-transfer interaction.  
Since pure momentum-transfer interactions leave the background cosmology unaffected, the background equations for all fluid components, including the interacting dark matter and baryons, retain their standard form:
\begin{equation}
\rho_{i} ' + 3 \mathcal{H}\left(1 +w_{i}\right)\rho_{i}=0\,,
\end{equation}
with $w_{i}=p_{i}/\rho_{i}$ the equation of state parameter for each component, and the Friedmann equation also remains the same:
\begin{equation}
\mathcal H^2 = \frac{8 \pi G a^2}{3 }\sum_i \rho_{i}\,,
\end{equation}
with $\mH=a'/a$ the Hubble expansion rate.

We are interested in the linear perturbations and, in particular, the Euler equations where the dark matter-baryon interaction enters.  
The perturbed line element for scalar perturbations in the synchronous gauge is given by\footnote{The relevant equations in the Newtonian gauge are displayed in the Appendix~\ref{app:new_eqs}.}
\begin{equation}
ds^2=a^2(\tau)\left[-d\tau^2+\left(\delta_{ij}+h_{ij}\right)dx^i dx^j\right],
\end{equation}
where, in Fourier space, the scalar mode is written as
$h_{ij}(\vec{k},\tau)=\hat{k}_i\hat{k}_j h(\vec{k},\tau)
+(\hat{k}_i\hat{k}_j-\delta_{ij}/3)6\eta(\vec{k},\tau)$, with $\vec{k}=k\hat{k}$, 
and $h$ and $\eta$ the two scalar metric perturbation variables. 

For the matter perturbation equations, we only introduce a coupling between dark matter and baryons, then the equations for photons and neutrinos do not change.\footnote{In a more general $w$CDM scenario, the perturbation equations for dark energy would also remain unchanged.}  
By analogy with Thomson scattering between baryons and photons, we follow a similar structure for the dark matter-baryon interaction. In particular, we propose an interaction term proportional to the relative velocity $(\theta_{\rm b}-\theta_{\rm dm})$, as follows\footnote{A way of covariantise this interaction would be having a coupling $Q^\mu= \gamma \rho_b^2 (u_{dm}^\mu - u_{b}^\mu)$  such that $\nabla T_{\rm b}^{\mu\nu}=Q^\mu=-\nabla T_{\rm c}^{\mu\nu}$ where we can normalise the coupling constant as $\gamma \to \frac{\gamma}{\Omega_{\rm b} \rho_{\rm cr}}$.
}
\begin{eqnarray}
\deltab'&=&-\left(\thetab +\frac{1}{2}h'\right)\,,\\
\thetab'&=&-\mathcal{H} \thetab+ c_s^2k^2\deltab
+{{R_{\rm T}}} \Gamma_{\rm T}(\theta_{{\rm ph}-\thetab})+  \Gamma_{\rm m}(\theta_{\rm dm} - \theta_{\rm b})\,, \label{eq:theta_b}\\ 
\deltadm'&=& -\left(\thetadm +\frac{1}{2}h'\right)\,,\\
\thetadm'&=&-\mathcal{H} \thetadm 
- \Gamma_{\rm m}R_{m}(\theta_{\rm dm} -\theta_{\rm b})\,,
\label{eq:theta_dm}
\end{eqnarray}
where we have used the standard notation for the density contrast $\delta_i\equiv \delta \rho_i/\rho_i$ and for the Fourier space velocity perturbation  $\theta\equiv i \vec{k}\cdot\vec{v}$, and $c_s^2\equiv \delta P_{\rm b}/\delta\rho_b$  is the squared effective sound speed of the baryonic fluid coupled to the photons
before recombination via Thomson scattering. The quantities $\tau_{\rm m}\equiv 1/\Gamma_{\rm m}$ and $R_{\rm m}$ are the equivalents of the mean conformal time between photon collisions with electrons $ \tau_{\rm e}=(an_{\rm e}\sigma_{\rm T})^{-1}$ and of the ratio between coupled fluids of Thomson scattering $R_{\rm T}\equiv3\rho_{b}/(4\rho_{\rm ph})$. 
So for the pure momentum transfer between dark matter and baryons they are defined by analogy as
\begin{eqnarray}
     \tau_{\rm m}&\equiv&(an_{\rm m}\sigma_{\rm m})^{-1}\;,\\
     R_{\rm m}&\equiv&\frac{\rho_{\rm b}}{\rho_{\rm dm}}\;.
\end{eqnarray}
In previous  pure momentum transfer models (see \cite{Asghari:2019qld,Figueruelo:2021elm,BeltranJimenez:2020iyx}), it is common to use the effective interaction rate $\Gamma$ and the ratio of the energy densities of the interacting fluids multiplied by the effective interaction rate  $\Gamma R$, which are just related to our previous quantities as follows:
\begin{eqnarray}
    \Gamma_{\rm m}&=&\frac{1}{\tau_{\rm m}}=\gamma a^{-2} \,,\\
    \Gamma_{\rm m} R_{\rm m}&=& \frac{R_{\rm m}}{\tau_{\rm m}}= \gamma a^{-2} \frac{\Omega_{\rm b}}{\Omega_{\rm dm}}\,,
\end{eqnarray}
where all constant quantities are absorbed into  $\gamma$.\footnote{We use $\gamma$ for consistency with our previous models, whose coupling parameters are denoted by $\alpha$ and $\beta$. Although this notation is unrelated to the conventional use of $\gamma$ to denote photons, it is perhaps not entirely inappropriate, as the photon fluid is ultimately responsible for the effective pressure exerted on the coupled dark matter--baryon fluid through Thomson scattering. To avoid any ambiguity, all quantities associated with photons are explicitly denoted by the subscript `ph'.} 
Therefore, $\gamma$ will be our only new parameter controlling the proposed pure momentum coupling.

\subsection{Tight-Coupling Approximation}
\label{subsec:TCA}
Before recombination, baryons and photons are tightly coupled through Thomson scattering, which induces efficient momentum exchange between the two fluids. This effect is already captured in equation~\eqref{eq:theta_b} by the third term $ \Gamma_{\rm T} (\theta_{\rm ph} - \theta_{\rm b})$, where $\Gamma_{\rm T}$ is the Thomson scattering baryon–photon coupling defined as
\begin{equation}
    \Gamma_{\rm T} \equiv \frac{4 \rho_{\rm ph}}{3 \rho_{\rm b}}\, a n_e \sigma_{\rm T}=R_{\rm T} \tau_c^{-1}\;.
\end{equation}
This definition is nothing more than the effective relative density of the interacting components, $R_{\rm T} \equiv 4 \rho_{\rm ph}/(3 \rho_{\rm b})$, 
multiplied by the effective interaction rate, $\tau_c^{-1} \equiv  a n_e \sigma_{\rm T}$,  where $\tau_c^{-1}$ is the inverse Thomson scattering time, and $n_e$ and $\sigma_{\rm T}$ the abundance of free electrons and  the Thomson scattering cross-section, respectively. 
 A similar term also appears in the corresponding equations for photons. 
 Considering the regime before recombination, the photon mean free path is much shorter than both the Hubble time ($\tau_c/\tau\ll 1$) and the wavelengths of cosmological perturbations ($k\tau_c\ll 1$). As a result, baryons and photons are forced to share a common bulk velocity to very high accuracy, while their relative velocity and higher photon multipoles are strongly suppressed. This tight coupling is responsible for the acoustic oscillations of the pre-recombination plasma. 
 
From a numerical perspective and in a standard scenario, the strong coupling between baryons and photons leads to stiff evolution equations due to the presence of terms proportional to the inverse Thomson scattering time $\tau_c^{-1}$. 
A direct integration of the full Boltzmann hierarchy in this regime would require prohibitively small time steps. This situation is well-known, and thus, Boltzmann solvers introduce the familiar Tight-Coupling Approximation (TCA) schemes. The TCA scheme provides an efficient alternative by exploiting the smallness of $\tau_c$ at early times and expanding the equations perturbatively in powers of $\tau_c$. Within this framework, it is customary to define the quantity $\Theta_{{\rm ph}-{\rm b}} \equiv \theta_{\rm ph} - \theta_{\rm b}$, which represents the difference between the velocity divergences of photons and baryons; 
its time derivative, $\Theta_{{\rm ph}-{\rm b} }'$, is commonly referred to as the baryon-photon slip. Along with the photon shear and higher-order multipoles, these quantities are considered small and are therefore expressed algebraically in terms of the density and velocity perturbations of the coupled fluid. This general TCA scheme is implemented in cosmological Boltzmann solvers to kick in prior to recombination
(see Ref.~\cite{Cyr-Racine:2010qdb} and \cite{2011JCAP...07..034B} for the full derivation of all the schemes used in \texttt{CAMB} and \texttt{CLASS}, respectively, in the absence of interaction).

In the presence of the baryon-dark matter interaction, a new term is introduced to the baryons Euler equation and then the expressions for $\Theta_{{\rm ph}-{\rm b} }$ and the baryon-photon slip are consequently modified. At leading order in the TCA, baryons and photons behave as a single effective fluid ($\theta_{\rm ph} = \theta_{\rm b}$) with sound speed $c_{{\rm ph}- {\rm b}}$. At next-to-leading order, $\Theta_{{\rm ph}-{\rm b} }$ and the baryon-photon slip are given by
\begin{align}
\Theta_{{\rm ph}-{\rm b}}&= - \frac{\tau_c}{1+R_{\rm T}}
\left[-\mathcal{H}\,\theta_{\rm b}+ k^2\!\left(c_s^2\,\delta_{\rm b}- \frac{1}{4}\delta_{\rm ph}+ \sigma_{\rm ph}\right)\right]+ \mathcal{O}(\tau_c^2)\,, \label{eq:slip}\\
\Theta_{{\rm ph}-{\rm b}}'&= \left( \frac{\dot{\tau}_c}{\tau_c} - \frac{2 \mathcal{H}}{1+R_T} + \frac{\Gamma_{\rm m}R_T}{1+R_T} \right) (\theta_{\rm ph}-\theta_{\rm b})
- \frac{\tau_c}{1+R_T} \left[- (\dot{\mathcal{H}}+ \mathcal{H}^2) \thetab 
- k^2 c_s^2 \Gamma_{\rm m}\delta_{\rm b}
- \frac{k^2}{2}\mathcal{H}\delta_{\rm ph} 
  +  k^2 \left(c_s^2 \dot{\delta}_{\rm b}  - \frac{1}{4}\dot{\delta}_{\rm ph}\right )
  \right. 
  \nonumber \\
 &
 \hspace{7.35cm} \left. 
  + (\thetadm-\thetab)\left(\dot{\Gamma}_{\rm m} - \Gamma_{\rm m}^2(1+R_{\rm m})\right)
  \right ] + \mathcal{O}(\tau_c^2)\,, \label{eq:slip_prime}
\end{align}
and the relevant equations for the TCA approximation are modified as  follow: 
\begin{eqnarray}
	\theta_{\rm b}' &=& \frac{1}{1+R_T}\left[
	-\mathcal{H}\,\theta_{\rm b}
		+ k^2\left( c_s^2 \delta_{\rm b} + R\left(\frac{\delta_{\rm ph}}{4} - s_2^2\,\sigma_{\rm ph}\right)
			\right) + \Gamma\left(\theta_{\rm dm} - \theta_{\rm b}\right)\right] 
			- \frac{R_T}{1+R_T} \Theta_{{\rm ph}-{\rm b}}' \,, \\
		\theta'_{\rm ph}& =&
		\frac{1}{R_T} \left[ -\theta'_{\rm b}- \mathcal{H}\,\theta_{\rm b}
			+ k^2 c_s^2 \delta_{\rm b} + \Gamma\,(\theta_{dm}-\theta_{\rm b})
		\right]
		+ k^2\left(\frac{\delta_{\rm ph}}{4} - \sigma_{\rm ph}\right)\,.
	\end{eqnarray}

Having derived the modified evolution equations of the model, in the next section we proceed to implement them numerically in Boltzmann codes. We first analyse the main effects of the interaction on cosmological observables and then confront the model with current observational data to derive constraints on the model parameter $\gamma$.    

\section{Numerical analysis}
\label{sec:numerical}
In the previous section, we derived the main modifications that are needed in the Euler equations~\eqref{eq:theta_b} and~\eqref{eq:theta_dm} due to the pure momentum transfer between dark matter and baryons. We also propagate those modifications to the TCA schemes usually used in the Boltzmann solvers in Section~\ref{subsec:TCA}. Therefore, we have modified \texttt{CAMB}~\cite{Lewis:1999bs} 
 and \texttt{CLASS}~\cite{2011arXiv1104.2932L,2011JCAP...07..034B} codes to incorporate all these changes\footnote{In particular,  
 for \texttt{CAMB} we have modified the so-called \lstinline[language=bash]{TightCoupling} scheme, while for \texttt{CLASS} the \lstinline[language=bash]{first_order_CLASS} scheme.}. Beside the \texttt{CLASS}/\texttt{CAMB} implementations, we have also created a modified version of \texttt{SymBoltz}~\cite{Sletmoen:2025fro} incorporating the dark matter-baryons elastic coupling\footnote{\href{https://github.com/david-figuer/SymBoltz_momentum_transfer}{https://github.com/david-figuer/SymBoltz\_momentum\_transfer}}. \texttt{SymBoltz} is a Boltzmann solver which has no approximations and which does not rely on TCA or similar schemes, therefore allowing us to compare the results independently of the scheme we have used in \texttt{CLASS} and {CAMB}. We compare the results among the three codes and found no significant differences, as detailed in Appendix~\ref{app:SymBoltz}. 

In the following analysis, we fix the values of the main cosmological parameters in order to clearly identify and isolate the physical effects under investigation. Specifically, we set the Hubble constant to $H_0 = 67.81\,\mathrm{km\,s^{-1}\,Mpc^{-1}}$, the baryon density is fixed to $\Omega_{\rm b} h^2 = 0.0224$ while the cold dark to $\Omega_{\mathrm{dm}} h^2 = 0.120$, and we consider a cosmological constant. We set the initial amplitude of scalar perturbations to be $A_{\rm s}=2.1\cdot10^{-9}$ and the spectral index to $n_{\rm s}=0.966$. The other cosmological parameters we keep to their default value in the code. Then, in the following analyses, we only vary the coupling parameter $\gamma$. We have performed the analysis using both \texttt{CLASS} and \texttt{CAMB} in order to verify that our results are consistent across independent Boltzmann solvers. This comparison allows us to explicitly check that the physical effects we report are not code-dependent. In particular, this cross-validation ensures that the specific treatment of the tight-coupling approximation (TCA), which is different in both codes, does not introduce significant differences in the resulting observables. 

\begin{figure}
\centering
\begin{tabular}{cc}
\includegraphics[width=0.49\textwidth]{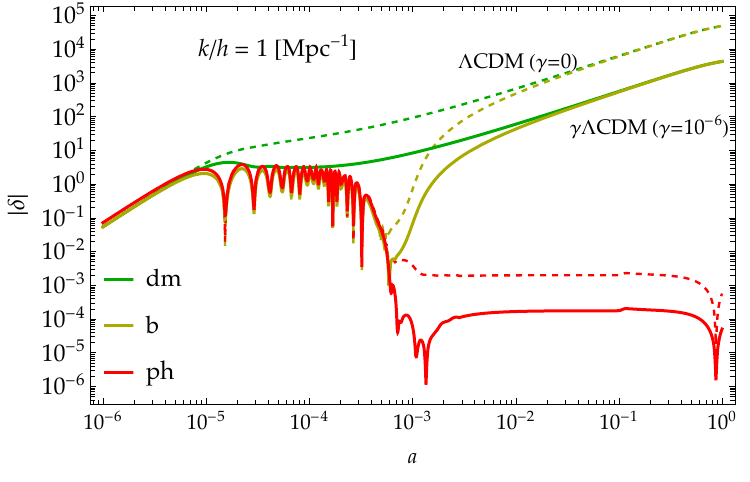}
\includegraphics[width=0.49\textwidth]{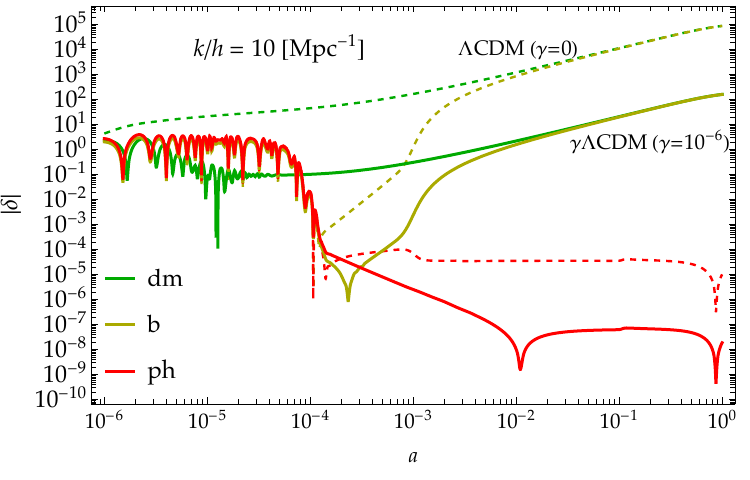}
\end{tabular}
\caption{Evolution of the density perturbations of CDM, baryons and photons components for the $k/h=1$Mpc$^{-1}$ (left) and  $k/h=10$Mpc$^{-1}$ (right)  scales. Dashed and solid lines correspond, respectively, to the fiducial $\Lambda$CDM model ($\gamma=0$) and the interacting $\gamma\Lambda$CDM model with $\gamma=10^{-6}$. }
\label{plotsDELTAS}
\end{figure}

We start by analysing the effect of the interaction on the evolution of the density perturbations for the dark-matter, baryon, and photon components. As already mentioned, the dark matter--baryon interaction is relevant at early times, before recombination, due to the effective  pressure of the tightly coupled baryon--photon plasma. We therefore expect smaller scales, which enter the Hubble horizon earlier, to be more sensitive to the effect of $\gamma$. This is indeed the behaviour shown in Figure~\ref{plotsDELTAS}, where we present the evolution of the density perturbations, $\delta_i$, $i=$dm,b,ph for the modes $k/h=1$Mpc$^{-1}$ (left) and $k/h=10$Mpc$^{-1}$ (right), assuming  $\gamma=10^{-6}$. In both cases,  the baryon-photon plasma slows the growth of $\delta_{\rm dm}$ before recombination and, as a consequence, the overall evolution of the matter perturbations is suppressed with respect to the standard $\Lambda$CDM case. Moreover, on sufficiently small scales ($k/h\sim10$Mpc$^{-1}$), $\delta_c$ also undergoes acoustic oscillations, which are completely absent in the standard scenario.

\begin{figure}
\centering
\begin{tabular}{cc}
\includegraphics[width=0.49\textwidth]{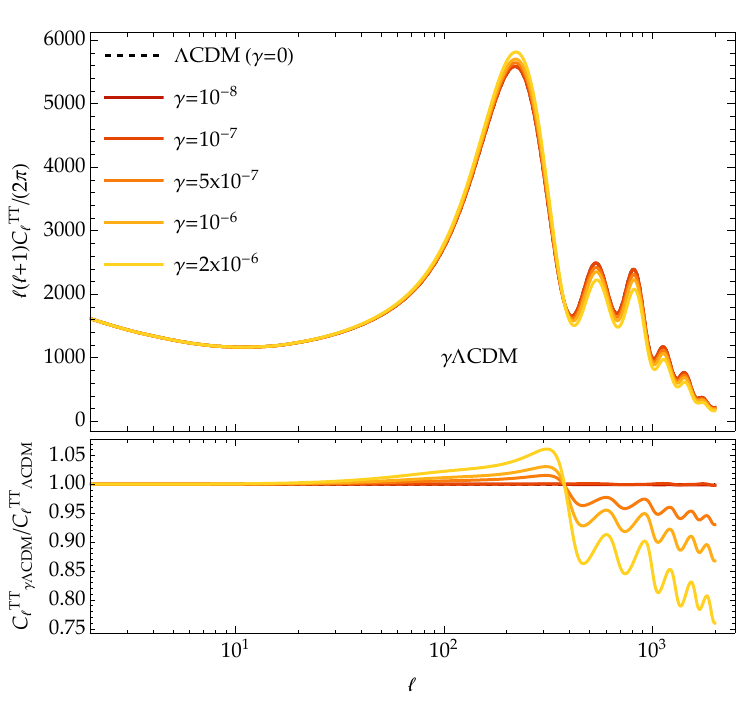}
\includegraphics[width=0.49\textwidth]{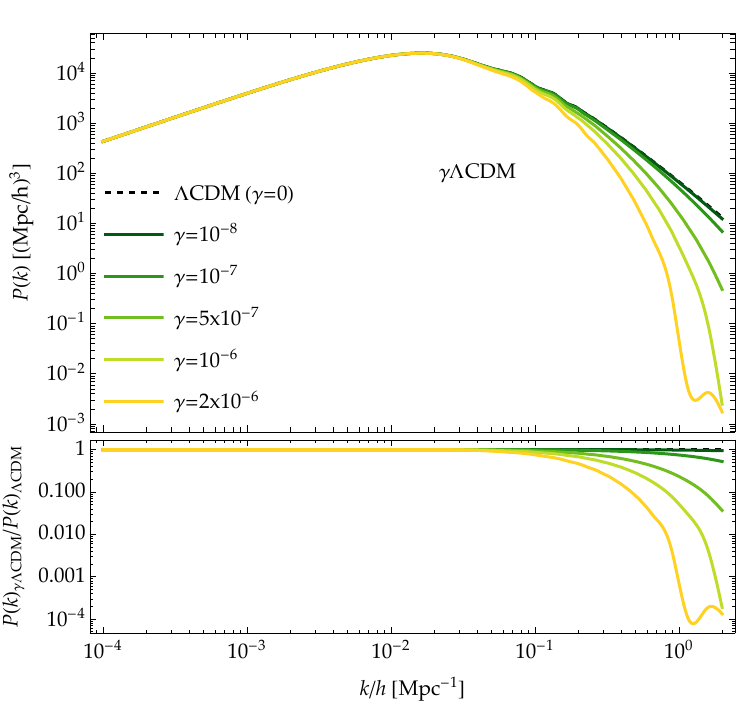}
\end{tabular}
\caption{Angular power spectrum for the CMB temperature anisotropies (left panel) and matter power spectrum (right panel) of the $\gamma\Lambda$CDM model for several values of the coupling parameter $\gamma$ within the range $[0, 10^{-5}]$. In both cases the relative errors are shown w.r.t. the fiducial $\Lambda$CDM.}
\label{plotsPkCLTT}
\end{figure}

Regarding the observational effects of the dark matter-baryons interaction, we first focus on the Cosmic Microwave Background (CMB) temperature power spectrum, shown in the left panel of Fig.~\ref{plotsPkCLTT} for several values of the coupling parameter $\gamma$ in the range $[0,10^{-6}]$. 
The additional dark matter-baryon interaction does not alter the total matter abundance. Instead, the drag experienced by dark matter effectively increases the inertia of the baryon component and, consequently, the photon--baryon momentum density ratio $R_T$. 
Physically, the interaction enhances gravitational compression and gravitational potentials decay more slowly.  
As a consequence, the impact of the $\gamma$-interaction on the CMB temperature power spectrum closely resembles that of increasing baryon loading. In particular, the amplitudes of the acoustic oscillations are enhanced, and the zero point of the oscillation is shifted, thereby modulating the relative heights of neighboring acoustic peaks: odd-numbered peaks are enhanced relative to the zero-baryon case, whereas even-numbered peaks remain nearly unchanged (see, for instance, \cite{WHu2003}). 
Overall, the deviation from the standard $\Lambda$CDM spectrum becomes more pronounced as the interaction strength $\gamma$ increases.

The effects of the interaction on the matter power spectrum are shown in the right panel of Fig.~\ref{plotsPkCLTT} for the same values of the coupling parameter $\gamma$. As $\gamma$ increases, we observe a greater suppression of power on small scales, while the position of the turnover scale remains essentially unchanged. This behaviour shares some similarities with the momentum transfer models between dark matter and dark energy studied in previous works~\cite{Figueruelo:2021elm}, but key differences also appear.  
As in the dark matter and dark energy pure momentum interacting scenarios, the suppression at small scales originates from an effective drag term in the Euler equations, which becomes relevant whenever there is a non-vanishing relative velocity between the interacting fluids, which for the present interaction is the early Universe, while for dark energy-dark matter and dark energy-baryons pure momentum coupling was the late Universe when dark energy emerges. Another key difference is the source of pressure that reduces the clustering. Here, both interacting species are pressureless, so it cannot be attributed to an intrinsic pressure or sound speed of either fluid in contrast to momentum transfer models involving dark energy, where the pressure of the dark energy component plays that role. However, here the physical origin of the suppression is instead associated with the tight coupling between baryons and photons prior to recombination. There, an effective pressure on baryons appears due to their coupling to photons, which then is transferred to dark matter and, hence, sources the pressure. Interestingly, these effects appear only for small scales, ($k\gtrsim 1$), as shown in Fig.~\ref{plotsDELTAS}. Those are the sub-horizon scales  at the recombination epoch, then sensitive to the effective pressure of the baryon-photon fluid during the Thomson scattering. Another important difference from the exchange momentum involving dark energy is the absence of a significant shift in the turnover scale. 
This can be understood also from the fact that the interaction becomes effective only for sub-horizon small-scale modes. Larger scales around the turnover remain essentially unaffected because baryons and dark matter share nearly the same velocity field there. 

As a final remark, it is worth noting that massive neutrinos are known to suppress the formation of structures on small scales, particularly after they become non-relativistic at low redshift. This raises the question of whether such an effect could be degenerate with that of the dark matter-baryon interaction, since both mechanisms produce a suppression in the matter power spectrum at small scales (see, for instance, Ref.~\cite{BeltranJimenez:2024lml}). Although suppression patterns on small scales exhibit some distinguishing features, their combined impact might lead to a degeneracy between the interaction parameter $\gamma$ and the neutrino mass $m_\nu$. This degeneracy would weaken the observational constraints on the interaction detection, an issue that will be explored in further detail in the next section.

\section{Observational constraints}
\label{sec:constraints}

Having explored the main phenomenological effects of the baryon-dark matter momentum transfer interaction, we now turn our attention to its compatibility with current cosmological observations. We perform a Markov Chain Monte Carlo analysis using the public code \texttt{Cobaya}~\cite{2021JCAP...05..057T,2019ascl.soft10019T} with our modified versions of \texttt{CAMB} and \texttt{CLASS}, finding no significant differences between the two implementations.

We consider the following dataset combinations for our MCMC analyses:
\begin{itemize}
    \item Baseline I (P18+PP+DESI): Planck PR4 likelihood \cite{Planck:2020olo} of temperature (TT), polarisation (EE), their cross-correlation (TE) \&  the Pantheon+ supernovae compilation (PP) \cite{Brout:2022vxf} \& the BAO measurements from DESI-DR2 \cite{DESI:2025zgx};
 \item Baseline II  (P18L+PP+DESI): same data as Baseline I but including the CMB lensing temperature+polarization-based from Planck 2018 \cite{Aghanim:2018oex};

\item Baseline I/II + $S_8$: we add a low-redshift constraint on the amplitude of matter fluctuations $S_8$. We use two different measurements to perform a robustness analysis of the obtained results:  $S_{8,{\rm SZ}} \equiv \sigma_8 (\Omega_{\rm m0}/0.27)^{0.3}
= 0.782 \pm 0.010$ derived from Sunyaev-Zeldovich (SZ) cluster counts~\cite{Planck:2013lkt} and $S_{8,{\rm DES}}\equiv (\Omega_{\rm m0}/0.3)^{0.5} = 0.789\pm0.012$ from DES year 6 results \cite{DES:2026fyc}. 
\end{itemize}
The distinction between Baselines I and II is motivated by the complementary information provided by CMB lensing on the growth of cosmic structures. Since the lensing signal probes the integrated matter distribution, it provides an additional handle on the evolution of matter perturbations beyond that encoded in the primary CMB anisotropies, making it particularly relevant for testing momentum-transfer interaction models. 
The measurement of $S_{8,\rm obs}$, on the other hand, is added as a Gaussian likelihood of the following form: 
$\log {\cal L}_{S_8} =
-\left(S_{8,{\rm th}}-S_{8,{\rm obs}}\right)^2/(2\sigma_{S_8}^2)$, 
where $S_{8,{\rm th}}$ is the theoretical prediction and $\sigma_{S_8}$ is the uncertainty of the measurement considered. 

 For our analysis, we vary the standard set of cosmological parameters consisting of the comoving sound horizon at
recombination as $100\theta_{\rm s}$, the today's density of baryons $\Omega_{\rm b} h^2$ and of cold dark matter $\Omega_{\mathrm{dm}} h^2$, the scalar spectral index $n_{\rm s}$ and the logarithm of the primordial amplitude $\ln(10^{10}A_{\rm s})$, and the optical depth to reionisation $\tau_{\mathrm{reio}}$.  As derived parameters, we compute the redshift of reionisation $z_{\mathrm{reio}}$, the total matter density parameter $\Omega_{\rm m}$, the amplitude of matter fluctuations on $8\,h^{-1}\mathrm{Mpc}$ scales $\sigma_8$ and the Hubble parameter $H_0$. We assume a flat Universe with a cosmological constant and fix all remaining cosmological parameters to their default values in \texttt{CAMB}\footnote{And to the default values of \texttt{CLASS}. We find no difference between the constraints obtained in both codes so that here we only present the \texttt{CAMB} results.}. Apart from those parameters, in our analyses, we are going to consider the following cosmologies:
\begin{itemize}[noitemsep]
    \item $\Lambda$CDM: the concordance model $\Lambda$CDM with two massless neutrinos and one massive neutrino with fixed total  mass $\sum m_\nu=0.06\; \text{eV}$. 
    
    \item $\gamma\Lambda$CDM: the model of elastic coupling between dark matter and baryons with a cosmological constant $\Lambda$ in the dark sector and with two massless neutrinos and one massive neutrino with fixed total mass $\sum m_\nu=0.06\text{eV}$.

    \item $\nu$-$\Lambda$CDM: the concordance model $\Lambda$CDM with two massless neutrinos and one massive neutrino with their  total  mass $\sum m_\nu$ as a free parameter. 

    \item $\nu$-$\gamma\Lambda$CDM:  the dark matter-baryons elastic coupling model with a cosmological constant $\Lambda$ in the dark sector and with two massless neutrinos and one massive neutrino with their  total  mass $\sum m_\nu$ as free parameter. 
\end{itemize}
The priors for the additional model parameters are $\log_{10}\gamma \in [-10,-6]$ and $\sum m_\nu \in [0,5]$. 
In order to assess the convergence of the chains, we use the Gelman-Rubin criteria~\cite{1992StaSc...7..457G}, which sets the analysis as converged when  $\vert R-1\vert < 0.01$. Then we have analysed the results with \texttt{GetDist}~\cite{Lewis:2019xzd}.

\subsection{Results}

\begin{figure}
\centering
\includegraphics[width=1\textwidth]{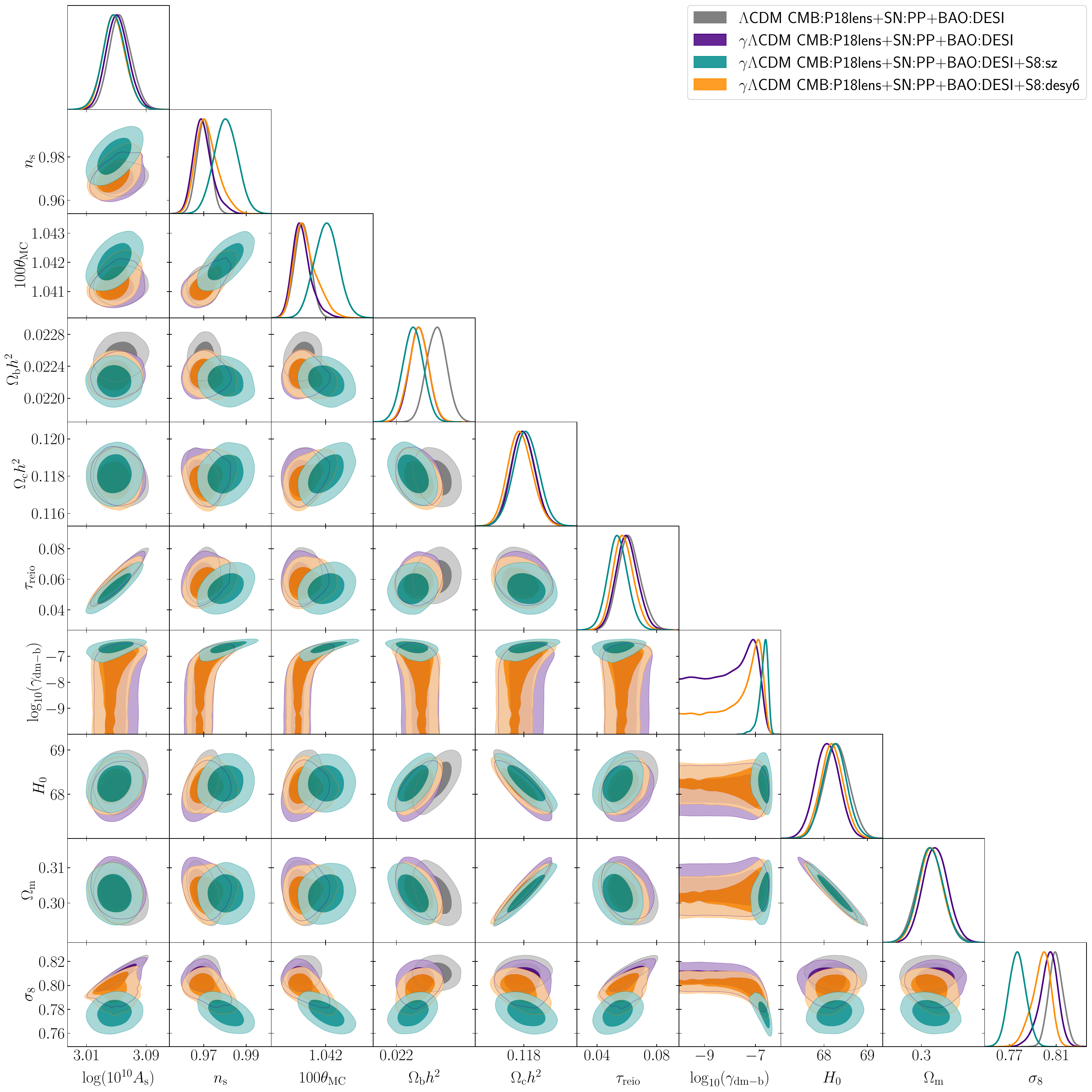}
\caption{Observational constraints on the parameters of the $\gamma\Lambda$CDM model using Baseline II dataset and the $S_8$ measurements from SZ and DES. The prior for the additional model parameter is $\log_{10}\gamma \in [-10,-6]$. }
\label{triangleGAMMA}
\end{figure}

We begin by analysing the MCMC results for the cosmologies $\Lambda$CDM and $\gamma\Lambda$CDM shown in Fig.~\ref{triangleGAMMA} and Table~\ref{tableGAMMA}, using the Baseline II and Baseline II + $S_8$ datasets. As expected for this kind of pure momentum interactions, the general parameter constraints in the  $\gamma\Lambda$CDM model are very similar to those obtained for the standard $\Lambda$CDM scenario. Nevertheless, the dark matter-baryon interaction is capable of reducing the preferred value of $\sigma_8$ while leaving the inferred value of $H_0$ essentially unchanged. Using only the Baseline II dataset, we obtain the upper limit $\log_{10}\gamma<-6.8$ (95\% C.L.) for the model parameter $\gamma$. The effect on clustering becomes more pronounced when low-redshift measurements of $S_8$ are included in the analysis.  In particular, the inclusion of $S_{8,{\rm SZ}}$ favours a non-zero interaction at more than the 2-$\sigma$ confidence level, yielding 
\begin{equation}
    \log_{10}\gamma=-6.73^{+0.26+0.39}_{-0.06-0.50}
\end{equation} 
in $68\,\%$ and $95\,\%$ C.L. This result is consistent with the behaviour previously found for the $\alpha$CDM \cite{Figueruelo:2021elm} and $\beta$CDM \cite{BeltranJimenez:2020iyx} models. The combination of Baseline II with the additional $S_{8,{\rm DES}}$ measurement exhibits the same trend, although in this case the preference for a non-vanishing interaction is not that strong and the $\Lambda$CDM scenario is not excluded.  

\begin{figure}
\centering
\includegraphics[width=1\textwidth]{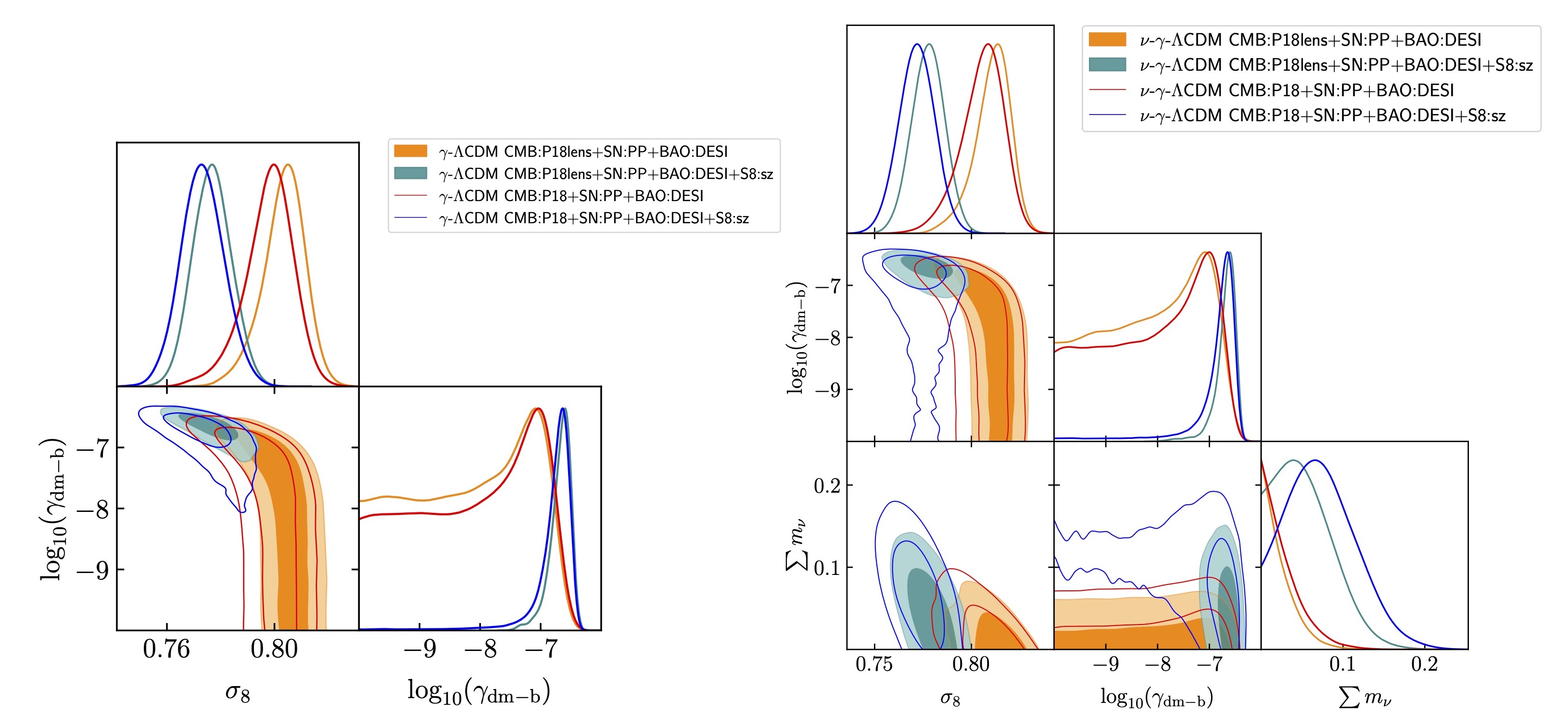}
\caption{Comparison of the observational constraints for $\gamma\Lambda$CDM and $\nu$-$\gamma\Lambda$CDM models between the Baseline I and Baseline II datasets. The addition of CMB lensing data from Planck 2018 \cite{Aghanim:2018oex} provides stronger evidence in favour of a non-zero interaction, driving the posterior distribution toward larger values of $\gamma$ and increasing the statistical significance of the interacting parameter being different from zero. }
\label{triangleLENS}
\end{figure}

In order to explore the impact of adding the Planck 2018 CMB temperature and polarization lensing data in our analysis, we show in the left panel of Figure~\ref{triangleLENS} the $\log_{10}\gamma$--$\sigma_8$ plane constraints obtained for the $\gamma\Lambda$CDM model using Baseline I and Baseline II. We can observe that the inclusion of the CMB lensing data noticeably strengthens the preference for a non-vanishing interaction, shifting the posterior towards larger values of $\gamma$ and increasing the statistical significance of the non-zero interacting parameter $\gamma$. 
This is indeed expected since CMB lensing probes the integrated matter distribution and the growth of large-scale structure between the last-scattering surface and the present epoch, thereby providing complementary information on the late-time evolution of matter perturbations. 

It is worth noting that, since the momentum transfer mediated by $\gamma$-interaction is effective at high redshifts, CMB observations constrain the model more tightly than the $\alpha$CDM and $\beta$CDM models, whose effects become significant primarily at low redshifts. 
Moreover, for both Baseline I and Baseline II datasets, the inferred baryon density $\Omega_{\rm b}h^2$ is reduced when the $\gamma$-interaction is present. This is because, as mentioned in Sect~\ref{sec:numerical}, the effects of momentum transfer between baryons and dark matter on the CMB temperature power spectrum have an effect similar to an effective increase in baryon loading. Consequently, observations favour lower values of the baryon density to compensate for this effect.

\begin{figure}
\centering
\includegraphics[width=1\textwidth]{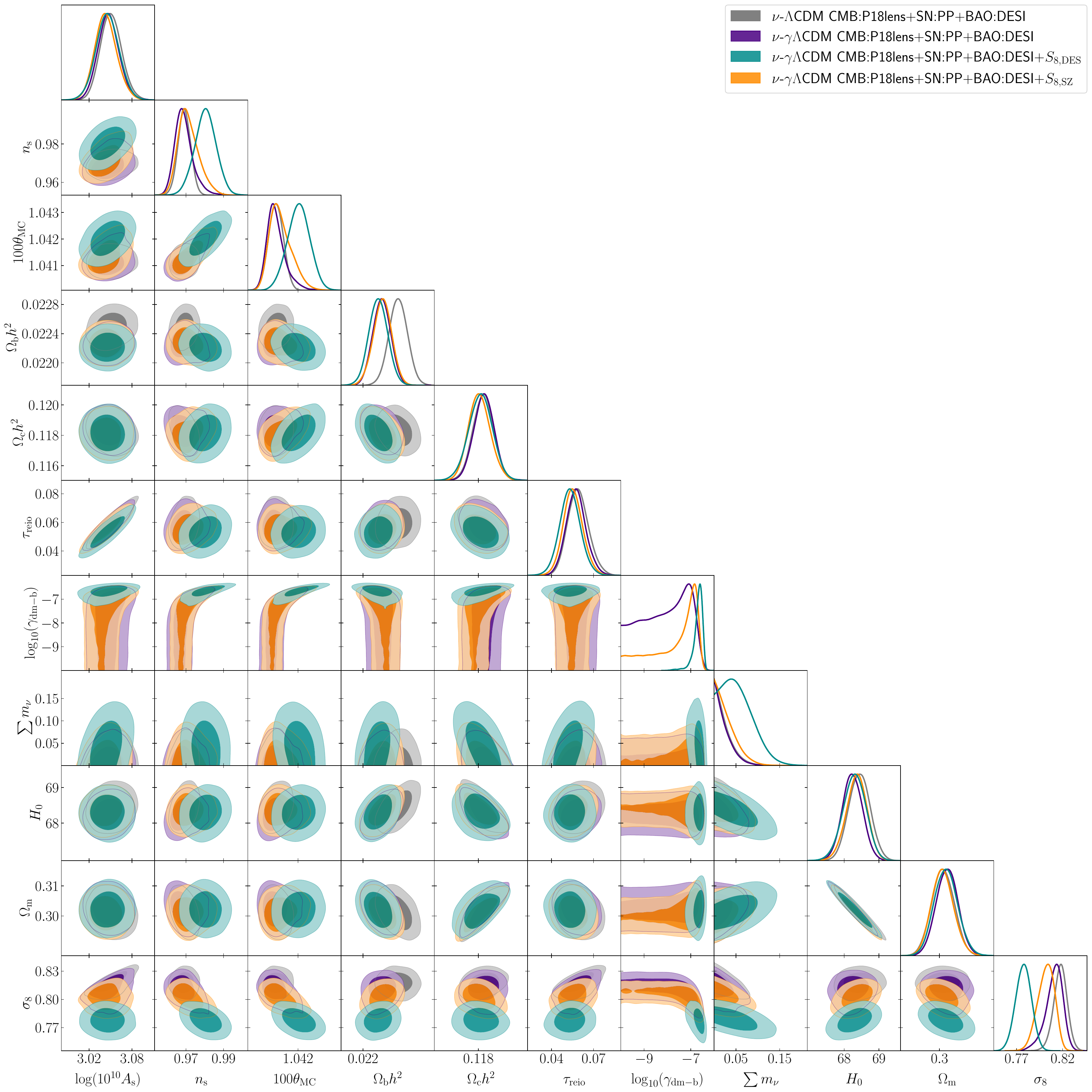}
\caption{Observational constraints on the parameters of the $\nu$-$\gamma\Lambda$CDM model using Baseline II dataset and the $S_8$ measurements from SZ and DES. The prior for the additional model parameters are $\log_{10}\gamma \in [-10,-6]$ and $\sum m_\nu \in [0,5]$. 
}
\label{triangleGAMMAmnu}
\end{figure}

\begin{figure}
\centering
\includegraphics[width=0.32\textwidth]{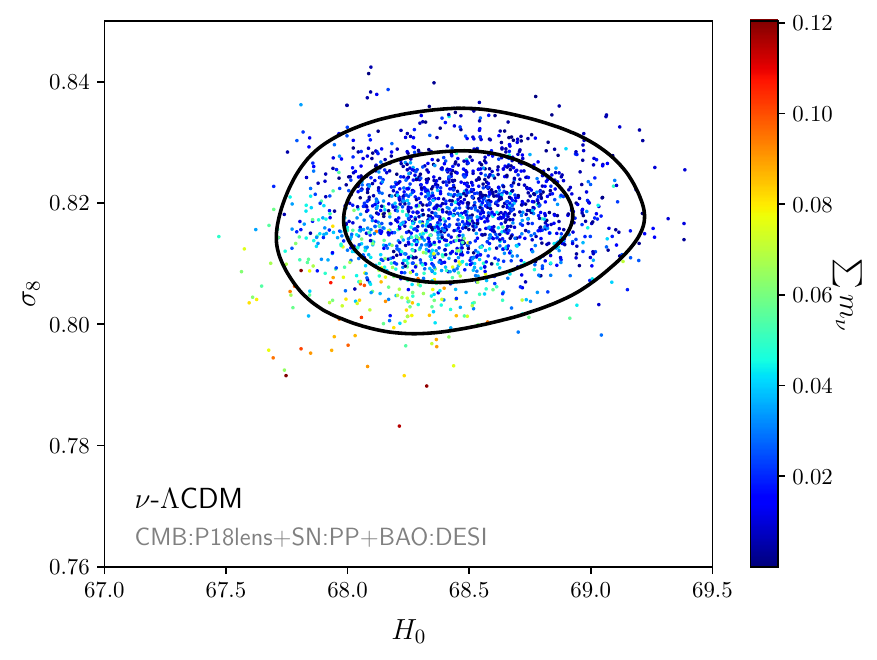} 
\includegraphics[width=0.32\textwidth]{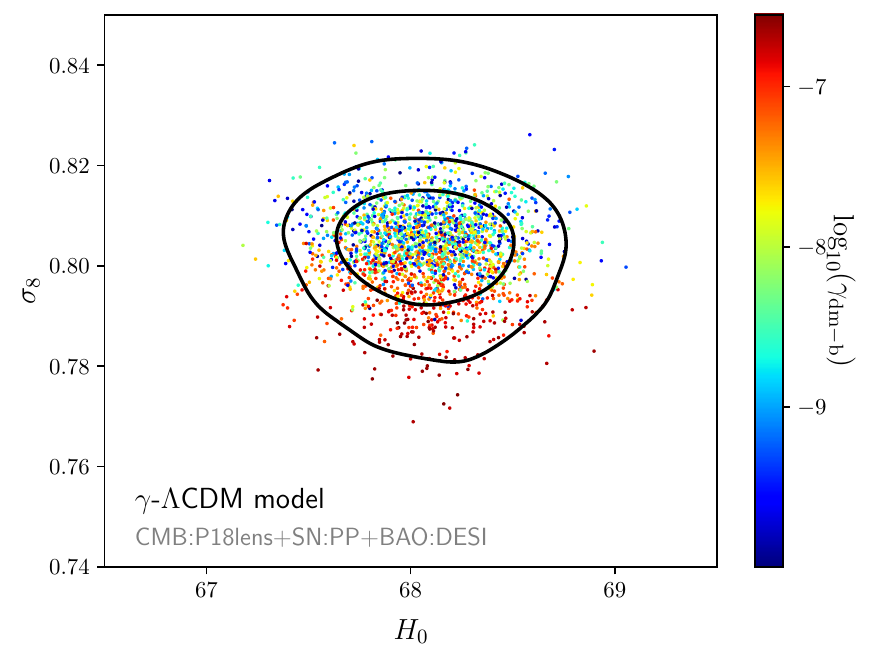} 
\includegraphics[width=0.32\textwidth]{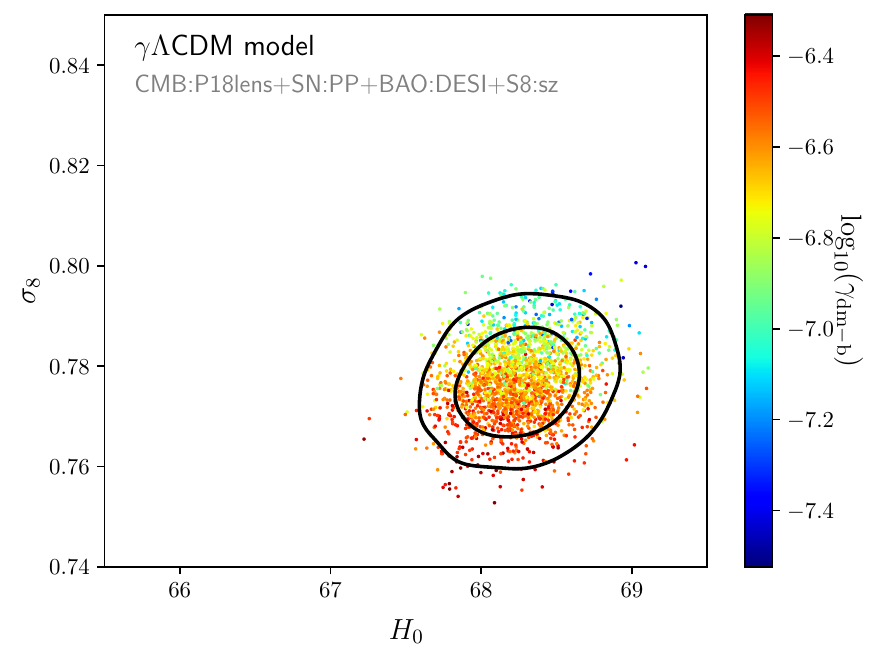} 
\\
\includegraphics[width=0.32\textwidth]{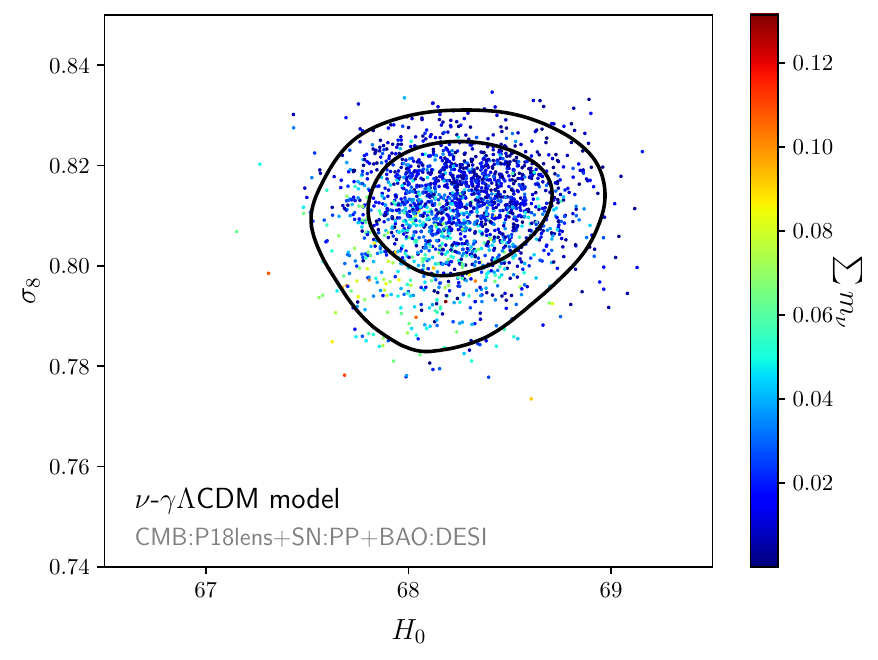}
\includegraphics[width=0.32\textwidth]{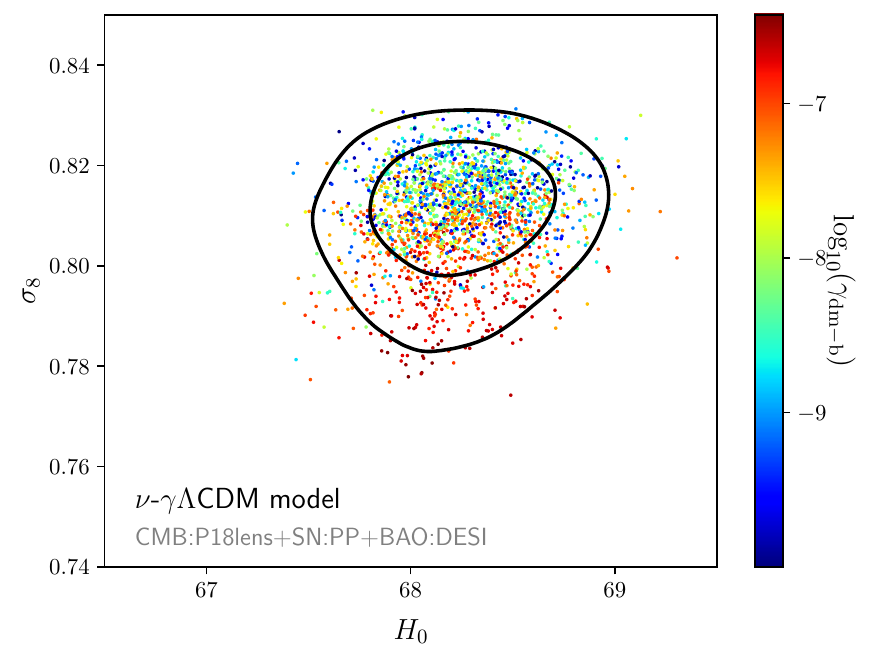}
\includegraphics[width=0.32\textwidth]{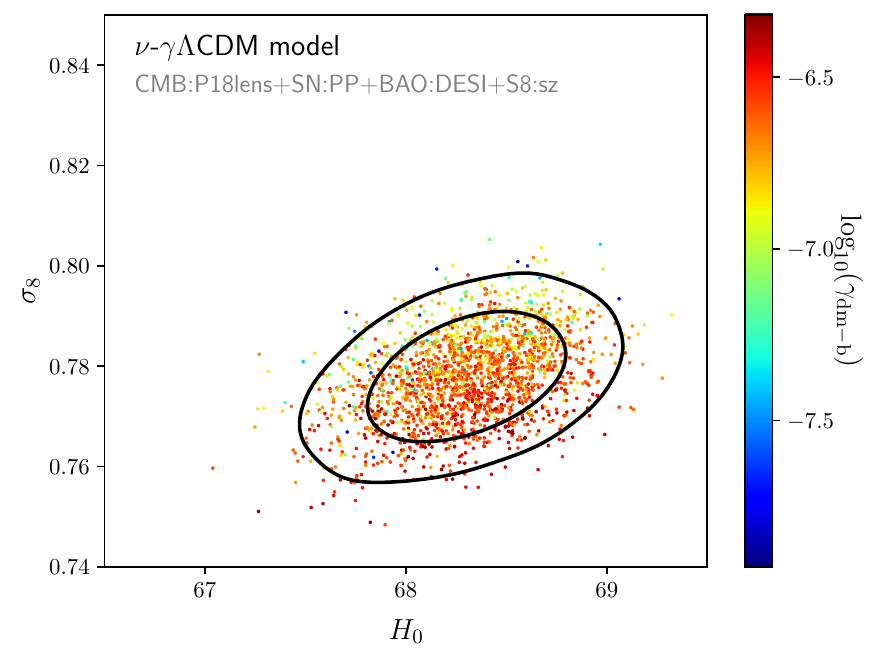}
\caption{In these set of panels, the $\sigma_8$--$H_0$ plane is shown coloured by the parameters $\log_{10}\gamma$ and $\sum m_\nu$ for the different scenarios considered using Baseline II dataset (+$S_{8,{\rm SZ}}$). }
\label{fig:pannel_sigma8_vs_H0}
\end{figure}

We now move towards the analysis of the observational constraints for the $\nu$-$\gamma\Lambda$CDM model, in which the sum of neutrino masses is allowed to vary. The results of the MCMC exploration are presented in Fig.~\ref{triangleGAMMAmnu} and Table~\ref{tableGAMMAmnu}. As in the previous case,  we can observe that the interaction reduces the preferred value of  $\sigma_8$  without significantly affecting the inferred value of $H_0$. Moreover, when the low-redshift measurement $S_{8,{\rm SZ}}$ is included, the data again favour a non-zero interaction: 
\begin{equation}
    \log_{10}\gamma=-6.70^{+0.24+0.36}_{-0.07-0.42}
\end{equation}
at $68\,\%$ and $95\,\%$ CL. 
The combined effect of the two parameters can be more clearly understood by analysing the $\sigma_8$--$H_0$ planes coloured by $\log_{10}\gamma$ and $\sum m_\nu$ for the different cosmological scenarios, as shown in Figure~\ref{fig:pannel_sigma8_vs_H0}. 
  In the first column, the mild degeneracy between $\sigma_8$, $H_0$ and $\sum m_\nu$ disappears in the $\nu$-$\gamma\Lambda$CDM scenario. That is, its residual effect in the usual $\nu$-$\Lambda$CDM model is completely removed by the presence of the interaction. In contrast, when comparing the $\gamma\Lambda$CDM and $\nu$-$\gamma\Lambda$CDM models (second column), the clear vertical color gradient in the $\sigma_8$--$H_0$ associated with the strength of the interaction remains essentially unchanged by the inclusion of massive neutrinos. This behavior suggests that neutrino mass does not significantly affect the inferred impact of the interaction parameter in the $\sigma_8–H_0$ plane, reinforcing the evidence that there is no significant degeneracy between the parameters $\gamma$ and $\sum m_\nu$. Moreover, once the $S_{8,{\rm S}Z}$ measurement is included (third column), the preferred values of $\sigma_8$ move down to the stronger interacting scenarios and the lowest values of $\gamma$ become completely excluded (that is, the statistical significance interaction increases at the 2-$\sigma$ w.r.t. the $\Lambda$CDM scenario).

Regarding the analysis using the Baseline I dataset (without including the CMB lensing data) for the $\nu$-$\gamma\Lambda$ CDM scenario, the observational constraints shown in the right panel of Figure~\ref{triangleLENS} show that the statistical significance of the interaction is reduced to approximately the level of 1-$\sigma$ when massive neutrinos are included. This weaker evidence for a non-zero interaction might be attributed to a possible degeneracy between $\gamma$ and the neutrino mass only when CMB lensing is not considered into the data analysis.  In fact, the $\log_{10}\gamma$--$\sum m_\nu$ panel reveals a clear asymmetry in the parameter degeneracy. For non-negligible values of the interaction strength $\gamma$, the preferred solutions remain concentrated around $\sum m_\nu \simeq 0$. By contrast, as the interaction strength decreases towards $\gamma\to 0$, the allowed parameter space extends towards non-zero neutrino masses, indicating that a non-zero $\sum m_\nu$ can be accommodated primarily in the limit where the interaction becomes negligible.

\begin{table}
\begin{center}
\renewcommand{\arraystretch}{1.8}
\begin{tabular}{ |c|c|c|c|c| } 
\hline
\hline
\centering
	 & $\Lambda$CDM Baseline II & $\gamma\Lambda$CDM Baseline II & $\gamma\Lambda$CDM Baseline II+$S_{8,\text{SZ}}$ & $\gamma\Lambda$CDM Baseline II+$S_{8,\text{DES}}$  \\
     \hline
	Parameter & mean\,$\pm\sigma\pm 2\sigma$ & mean\,$\pm\sigma\pm 2\sigma$  &  mean\,$\pm\sigma \pm 2\sigma$ & mean\,$\pm\sigma \pm 2\sigma$\\
	\hline \hline 
$100\Omega_{b0} h^2$ & $2.252\pm 0.013\pm 0.025 $ & $2.229\pm 0.012\pm 0.024 $ & $2.222\pm 0.013\pm0.026$ & $2.229\pm 0.013\pm-0.024$  \\\hline
$\Omega_{c0}h^2$  & $0.1179\pm 0.0007 \pm 0.0013$  & $0.118\pm 0.0007^{+0.0013}_{-0.0012}$ & $0.11808\pm 0.00072\pm 0.0014$  & $0.11779^{+0.00063+0.0014}_{-0.00071-0.0013}$ \\\hline 
$n_{s}$ & $0.9700\pm 0.0033 ^{+0.0064}_{-0.0065}$ & $0.9696^{+0.0035+0.0097}_{-0.0050-0.0086}$& $0.9802\pm 0.0059^{+0.011}_{-0.012} $  & $0.9716^{+0.0041+0.011}_{-0.0060-0.0095}$ \\ \hline  
$\log(10^{10}A_{s})$ & $3.055\pm 0.015 ^{+0.030}_{-0.028} $   & $3.052^{+0.014+0.031}_{-0.016-0.028}$ &  $3.047\pm 0.016 ^{+0.031}_{-0.032}$ & $3.048^{+0.014+0.031}_{-0.016-0.028}$ \\ \hline  
$\tau_{\rm reio}$  & $0.062^{+0.007+0.015}_{-0.008-0.014}$ & $0.0601^{+0.0067+0.015}_{-0.0076-0.014}$ & $0.0537\pm 0.0069\pm0.014 $   & $0.0577^{+0.0063+0.014}_{-0.0073-0.013}$   \\ \hline  
$100~\theta_{s}$  &  $1.0412\pm 0.0003 \pm 0.0005$ &  $1.0412^{+0.0002+0.0007}_{-0.0004-0.0006}$ & $1.042^{+0.0005+0.0008}_{-0.0004-0.0009}$ & $1.0413^{+0.0003+0.0008}_{-0.0005-0.0007}$ \\ \hline \hline
$H_0$\scriptsize{[km/s/Mpc]} & $68.28\pm 0.30 \pm 0.58 $  & $68.07\pm 0.28^{+0.54}_{-0.55}$ &  $68.25\pm 0.28^{+0.55}_{-0.53}$ &  $68.18\pm 0.26 \pm 0.52$ \\ \hline 
$\sigma_8$  & $0.81\pm 0.006 \pm 0.012 $  & $0.803^{+0.009+0.015}_{-0.006-0.017} $ &  $0.777^{+0.007+0.016}_{-0.008-0.014}$ & $0.797^{+0.009+0.014}_{-0.006-0.017}$ \\ \hline 
$\Omega_{m0}$  &  $0.303\pm 0.004 ^{+0.008}_{-0.007}$ & $0.304\pm 0.004\pm0.007$ & $0.303\pm 0.004\pm 0.007 $  & $0.303\pm 0.0035 \pm 0.007$ \\ \hline  
\hline
$\log_{10}\gamma$ & -  & $<-6.8^\ast$  & $-6.73^{+0.26+0.39}_{-0.06-0.50}$   & $<-6.67^\ast$ \\ \hline 
\hline
\end{tabular}
\end{center}
\caption{
MCMC results for the mean values and the $1\sigma$ ($68\,\%$ CL) and $2\sigma$ ($95\,\%$ CL) upper and lower limits of the cosmological and derived parameters for the $\Lambda$CDM and 
$\gamma\Lambda$CDM models using the Baseline II and Baseline II+$S_8$ datasets. The prior for the additional model parameter is $\log_{10}\gamma \in [-10,-6]$; the symbol  $^\ast$ indicates that the lower $95\%$ credible interval is prior-limited. 
}
\label{tableGAMMA}
\end{table}

\begin{table}
\begin{center}
\renewcommand{\arraystretch}{1.8}
\begin{tabular}{ |c|c|c|c|c| } 
\hline
\hline
\centering
	 &$\nu$-$\Lambda$CDM Baseline II & $\nu$-$\gamma\Lambda$CDM Baseline II & $\nu$-$\gamma\Lambda$CDM Baseline II+$S_{8,\text{SZ}}$  &  $\nu$-$\gamma\Lambda$CDM Baseline II+$S_{8,\text{DES}}$ \\ 
     \hline
	Parameter & mean\,$\pm\sigma\pm 2\sigma$ & mean\,$\pm\sigma\pm 2\sigma$  &  mean\,$\pm\sigma \pm 2\sigma$ & mean\,$\pm\sigma \pm 2\sigma$\\
	\hline \hline 
$100\Omega_{b0} h^2$ &$2.249\pm 0.013\pm 0.025$  & $2.226\pm 0.012^\pm 0.024$  & $2.221\pm 0.013 ^{+0.026}_{-0.025}$  &  $2.227\pm 0.013        ^{+0.025}_{-0.026}$    \\\hline
$\Omega_{c0}h^2$  & $0.1183\pm 0.0007 ^{+0.0013}_{-0.0014}$ & $0.1184\pm 0.0007\pm 0.0013$ & $0.1181\pm 0.0008 ^{+0.0015}_{-0.0016}$ & $0.1181^{+0.0007+0.0015}_{-0.0008-0.0014}$  \\\hline 
$n_{s}$ & $0.969\pm 0.003 \pm 0.007$ & $0.969^{+0.004+0.010}_{-0.005-0.009} $ & $0.980\pm 0.006 ^{+0.011}_{-0.012}$ &  $0.972^{+0.004+0.011}_{-0.006-0.010}$ \\ \hline  
$\log(10^{10}A_{s})$ & $3.051\pm 0.015   ^{+0.031}_{-0.028}  $ & $3.047\pm 0.015 ^{+0.030}_{-0.028} $ & $3.046\pm 0.016\pm 0.032$ & $3.044^{+0.014+0.032}_{-0.016-0.029} $ \\ \hline  
$\tau_{\rm reio}$  & $0.059^{+0.007+0.016}_{-0.008-0.014}$ & $0.058^{+0.007+0.015}_{-0.007-0.014}$ & $0.053\pm 0.007 ^{+0.015}_{-0.014}$   & $0.056\pm 0.007 ^{+0.015}_{-0.014} $   \\ \hline  
$100~\theta_{s}$  & $1.0412\pm 0.0003 ^{+0.0006}_{-0.0005}$  &  $1.0412^{+0.0002+0.0008}_{-0.0004-0.0006}$  & $1.0420\pm 0.0004 ^{+0.0008}_{-0.0009} $   & $1.0413^{+0.0003+0.0009}_{-0.0005-0.0007}$  \\ \hline \hline
$H_0$\scriptsize{[km/s/Mpc]} &  $68.44\pm 0.30  ^{+0.60}_{-0.59} $ & $68.24\pm 0.29 ^{+0.57}_{-0.57} $  & $68.30\pm 0.33 ^{+0.63}_{-0.66} $ &  $68.36^{+0.31+0.56}_{-0.28-0.60} $  \\ \hline 
$\sigma_8$  & $0.817^{+0.008+0.014}_{-0.007-0.016} $   & $0.811^{+0.011+0.017}_{-0.007-0.020} $ & $0.778\pm 0.009\pm0.017$  &  $0.802^{+0.010+0.017}_{-0.0079-0.018}$   \\ \hline  
$\Omega_{m0}$  & $0.301\pm 0.004 \pm 0.007$ & $0.303\pm 0.004 \pm 0.007$& $0.302\pm 0.004\pm 0.008$ & $0.301 \pm 0.004 \,{}^{+0.008}_{-0.007}$  \\ \hline  
\hline
$\log_{10}\gamma$ & - & $<-6.76^\ast$ &  $-6.70^{+0.24+0.36}_{-0.07-0.42} $ &  $<-6.64^\ast   $  \\ \hline 
$\sum m_\nu$ & $<0.066^\ast$  & $ <0.0623^\ast $ & 
$<0.118^\ast $&  
$<0.076^\ast$\\ \hline 
\hline
\end{tabular}
\end{center}
\caption{
MCMC results for the mean values and the $1\sigma$ ($68\,\%$ CL) and $2\sigma$ ($95\,\%$ CL) upper and lower limits of the cosmological and derived parameters for the $\nu$-$\Lambda$CDM and 
$\nu$-$\gamma\Lambda$CDM models using the Baseline II and Baseline II+$S_8$ datasets. The priors for the additional model parameters are $\log_{10}\gamma \in [-10,-6]$ and $\sum m_\nu \in [0,5]$; the symbol  $^\ast$ indicates that the lower $68\%$ (or $95\%$) credible interval is prior-limited. 
 }
\label{tableGAMMAmnu}
\end{table}

\section{Conclusions}
\label{sec:conclusions}

In this work, we have explored a cosmological model with pure momentum transfer between dark-matter and baryon fluids, namely $\gamma\Lambda$CDM. The model completes  the trilogy of interacting cosmological scenarios involving dark matter, dark energy and baryons. The interaction does not involve any energy exchange and affects the cosmological evolution exclusively at the level of the cosmological perturbations. 

We have derived the dynamical equations governing the model and implemented them consistently in the Boltzmann solvers \texttt{CAMB}, \texttt{CLASS}, and \texttt{SymBoltz}, finding excellent agreement among the three independent implementations. We  also modified the tight-coupling to consistently incorporate the interaction, ensuring numerical stability. Our analysis of the CMB temperature anisotropy and matter power spectra confirms that the $\gamma\Lambda$CDM model shares the characteristic feature previously identified in the  $\alpha$CDM and $\beta$CDM scenarios: a suppression of the matter power spectrum on small scales, leading to reduced late-time clustering as a consequence of the drag exerted by the pressure-supported fluid (radiation, in this case) on pressureless matter. 
The crucial difference here is that the effect of the $\gamma$-interaction on the observables originates at high redshifts, before recombination, whereas the interactions in the  two previously discussed models become relevant only at low redshifts.

We also obtained observational constraints on the model using combinations of  CMB, BAO, and SN datasets. For the case of the Baseline II dataset, we obtain the upper limit 
$\log_{10}\gamma<-6.8$ (95\% C.L.).  When the low-redshift $S_{8,SZ}$ measurement is included, the interacting scenario is preferred over the standard cosmological model, with a non-zero  interaction parameter detected at more than the 2-$\sigma$ level: $ \log_{10}\gamma=-6.70^{+0.24+0.36}_{-0.07-0.42}$ (68\% and 95\% C.L.). 
 Furthermore, we have investigated possible degeneracies between the effects of the dark matter-baryon interaction and massive neutrinos. We find that for the Baseline II dataset, both effects are distinguishable, although a weak degeneracy may emerge when CMB lensing information is excluded from the analysis. 

 In general, our results demonstrate that the $\gamma\Lambda$CDM scenario provides a viable extension of the standard cosmological model, preserving its background evolution while introducing distinctive signatures in the growth of cosmic structures. Future high-precision cosmological observations will allow us to further test this interaction and to determine whether pure momentum exchange between the dark sector and the baryonic component can play a role in resolving the remaining tensions in the standard cosmological paradigm.

\FloatBarrier
\begin{acknowledgments}
 The authors thank Jose Beltrán Jiménez for useful comments and insightful discussions that helped improve this work. 
We also thank Herman Sletmoen, the creator of the SymBoltz code,  for his explanations on how SymBoltz works and should be modified, and for his help in  the development of the SymBoltz implementation of this model. 
FATP thanks the hospitality of the Institute of Theoretical Astrophysics of the University of Oslo, where part of this work was performed. 
DF and FATP acknowledge the support of the grant PID2024-158938NB-I00 funded by MICIU/AEI/10.13039/ 501100011033 and by “ERDF A way of making Europe”. 
DF acknowledges support from "Convocatoria de contratación de    personal investigador doctor de la UPV/EHU (2024)" and 
 support from the Grant PID2021-123226NB-I00 (funded by MCIN/AEI / 10.13039/501100011033 and by “ERDF A way to make Europe”).
   FATP acknowledges the support of the {\it Programa de estancias de movilidad,  Modalidad Junior José Castillejo (2023)} by the Ministerio de Ciencia, Innovación y Universidades, Spain, and 
 the support of grants from Project SA097P24 funded by Junta de Castilla y Le\'on.
\end{acknowledgments}

\appendix

\section{Equations in the Newtonian gauge}
\label{app:new_eqs}

In the Newtonian gauge, the perturbed line element for scalar perturbations is given by
\begin{equation}
    \dd s^2 = a^2(\tau)\left[ (1+2\Phi) \tau^2 + (1-2\Psi)\delta_{ij}\dd x^i \dd x^j\right]\; ,
\end{equation}
 where $\tau$ is the conformal time, $x^i$ the comoving coordinates, $a$ the scale factor while $\Phi$ and $\Psi$ are the metric perturbations of the gauge. Proceeding as before, the interaction only modifies the perturbation regime and, in particular, the Euler equations as
\begin{eqnarray}
\label{eq:deltab}
\deltab'&=& -\thetab +3\Psi'  \;, \\
\label{eq:thetab}
\thetab'&=&-\mathcal{H} \thetadm + k^2 \Phi + c_s^2k^2\deltab
+{R_{\rm T}} \Gamma_{\rm T}(\theta_{\rm ph}-\thetab)+  \Gamma_{\rm m}(\theta_{\rm dm} - \theta_{\rm b})\;, \\
\label{eq:deltadm}
\deltadm'&=&-\thetadm + 3 \Psi' \;, \\
\label{eq:thetadm}
\thetadm'&=&-\mathcal{H} \thetadm + k^2 \Phi -\Gamma_{\rm m} R_{\rm m}(\thetadm-\thetab)\;, 
\end{eqnarray}
where, as before, $\Gamma_{\rm m}$ and $R_{\rm m}$ are the effective interaction rate and the relative density, respectively, given by\\
\begin{eqnarray}
    \Gamma_{\rm m}&=&\frac{1}{\tau_{\rm m}}=\gamma a^{-2} \,,\\
    \Gamma_{\rm m} R_{\rm m}&=& \frac{R_{\rm m}}{\tau_{\rm m}}= \gamma a^{-2} \frac{\Omega_{\rm b}}{\Omega_{\rm dm}}\,.
\end{eqnarray}

Provided the stiffness of the evolution equations for the coupled system of photons and baryons due the terms proportional to the inverse of the Thomson scattering time $\tau_c^{-1}$, \texttt{CLASS} Boltzmann solvers make use of the Tight-Coupling Approximation schemes (TCA).  \texttt{CLASS} code has several different options but, in our case, we are going to focus in the \lstinline[language=bash]{first_order_CLASS}  scheme as the reference one which we will modify to add the new interacting term between dark matter and baryons. In particular, the Euler equations for baryons and photons have to be rewritten as 
\begin{align}
    \thetab'&=\frac{1}{1+R_{\rm T}}\left[{-\mathcal{H}\theta_b+k^2\left(c_s^2 \deltab+R\left(\frac{\delta_{\rm ph}}{4}- \sigma_{\rm ph}\right)\right)+ \Gamma\left(\theta_{\mathrm{dm}}-\theta_b\right)- R_{\rm T} \Theta_{{\rm ph}-{\rm b}}'}\right]+k^2 \Phi      \;,\\
    \theta_{\rm ph}'&= k^2 \left[  \frac{\delta_{\rm ph}}{4} - \sigma_{\rm ph} \right] + k^2 \Phi - \frac{1}{R}\left[\thetab' +\mathcal{H} \thetab -c_s^2 k^2 \deltab - k^2 \Phi - \Gamma(\theta_{\rm dm} -\thetab) \right]\;.
\end{align}
The equation for $\Theta_{{\rm ph}-{\rm b}}$ and  $\Theta_{{\rm ph}-{\rm b}}'$ are construct from subtracting eq~\eqref{eq:theta_b} for $\thetab$ and the corresponding equation for $\theta_{\rm ph}$, which leads to a gauge independent equation and therefore the same equations for $\Theta_{{\rm ph}-{\rm b}}$ and $\Theta_{{\rm ph}-{\rm b}}'$ as the ones in eq.~\eqref{eq:slip} and eq.~\eqref{eq:slip_prime}  apply.

\section{Validation of \texttt{SymBoltz} code}
\label{app:SymBoltz}
As commented before, we have also created an implementation of the studied coupling in the new \texttt{SymBoltz} Einstein-Boltzmann solver~\cite{Sletmoen:2025fro}. \texttt{SymBoltz} is a recent Julia package to solve the linear Einstein-Boltzmann equations. In contrast to standard Boltzmann solvers like \texttt{CLASS} or \texttt{CAMB}, it does not rely on approximation-switching schemes, since it integrates the full stiff system at all times by means of implicit methods. That code provides a symbolic-numeric interface that allows the user to specify the relevant equations in a flexible and modular way and, in addition, it is fully compatible with automatic differentiation, which can be particularly profitable for forecast analyses. In our case, we have created a version of \texttt{SymBoltz} that performs all the pure momentum coupling studied by the authors of this paper. In particular, we have included the model studied here performing a pure momentum coupling between dark matter and baryons here studied, but also the pure momentum transfer between dark energy and dark matter ($\alpha$CDM) analysed in Ref.~\cite{Figueruelo:2021elm} and between dark energy and baryons ($\beta$CDM) studied in Ref.~\cite{BeltranJimenez:2020iyx}. The repository with the code  can be found at \href{https://github.com/david-figuer/SymBoltz_momentum_transfer}{https://github.com/david-figuer/SymBoltz\_momentum\_transfer}.\\

\begin{figure}
\centering
\includegraphics[width=0.49\textwidth]{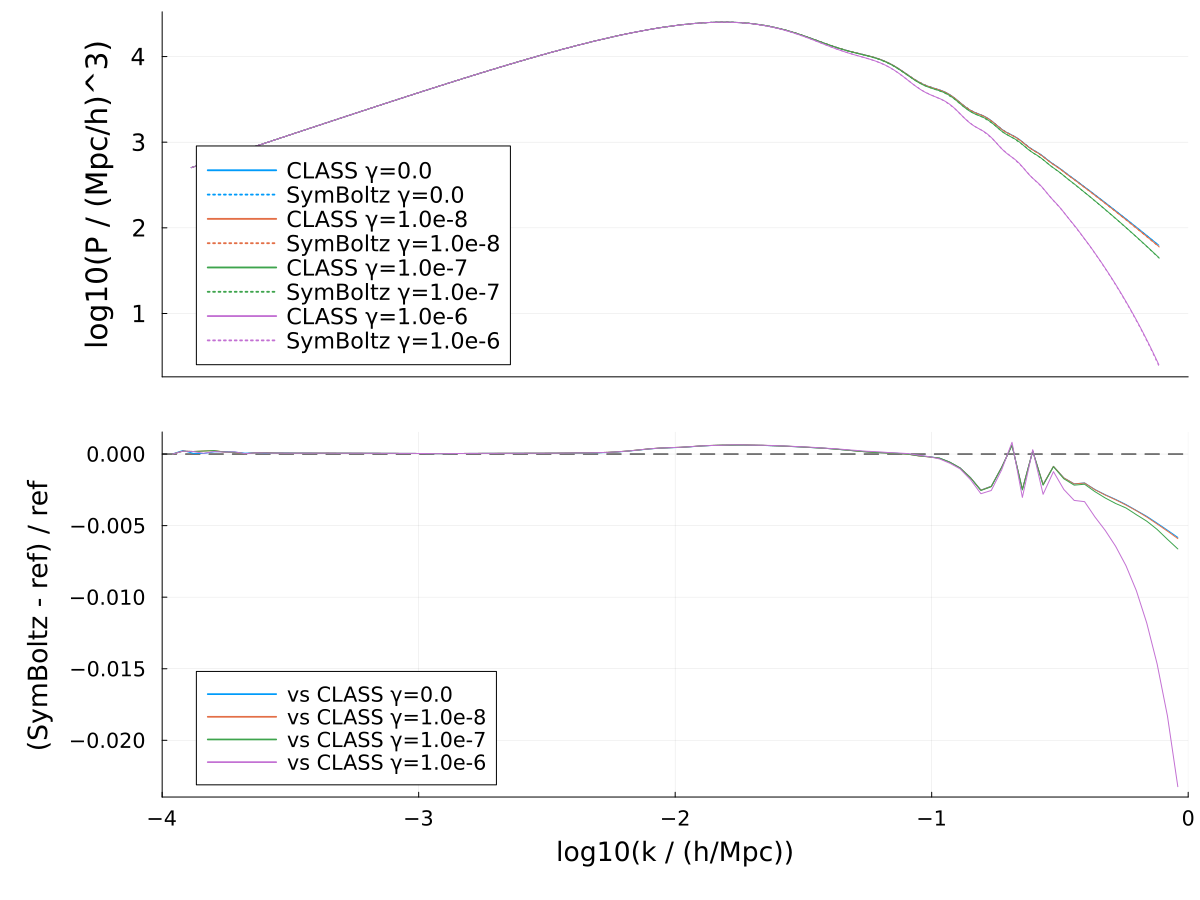}
\includegraphics[width=0.49\textwidth]{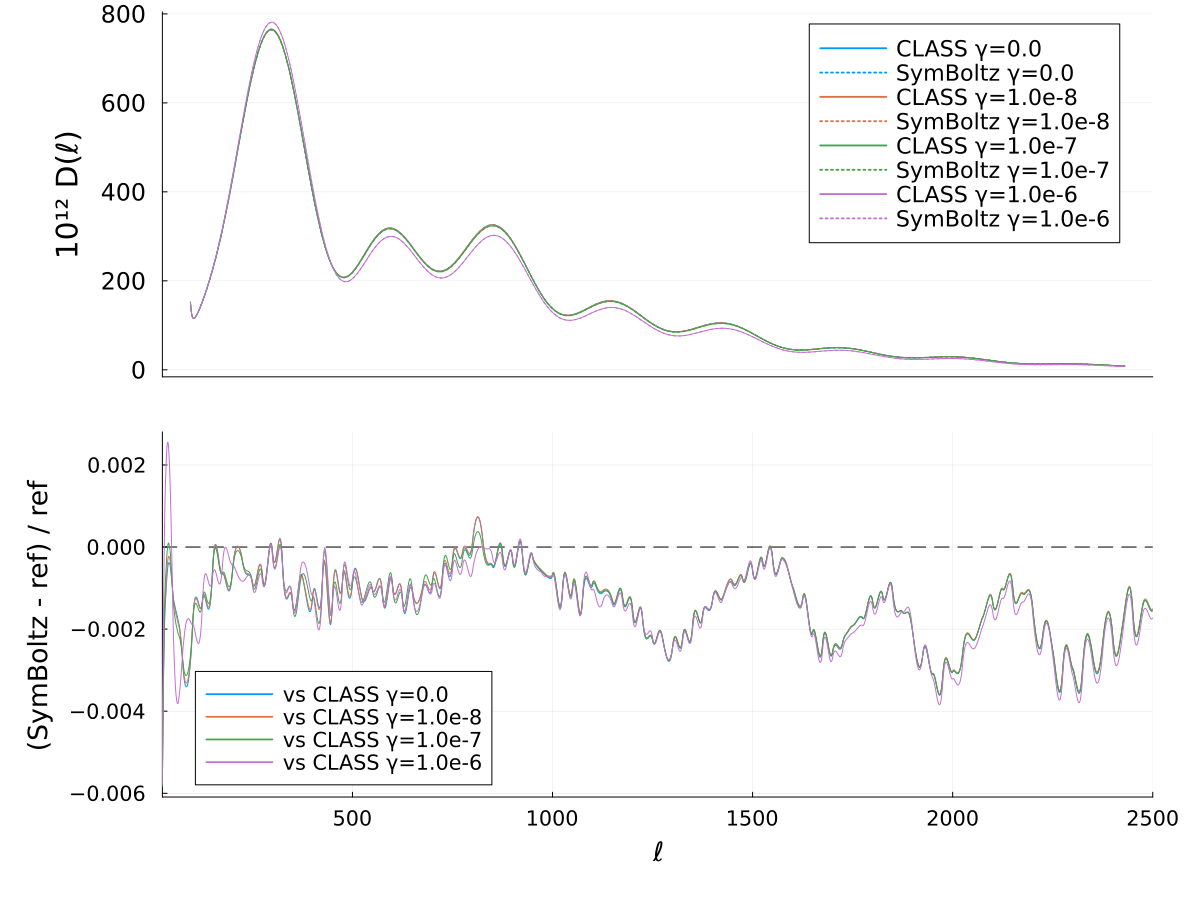}
\caption{\texttt{SymBoltz} implementation vs \texttt{CLASS} implementation: in the left plot we display the matter power spectrum for several values of the coupling parameter $\gamma$ while in the right plot we display the temperature autocorrelation of the cosmic microwave background. \texttt{SymBoltz} and \texttt{CLASS} show compatible results, even though \texttt{CLASS} require a \lstinline[language=bash]{first_order_CLASS} approximation scheme for the tight-coupled system formed by baryons and photons while \texttt{SymBoltz}  is free from any approximation.}
\label{fig:validation_SymBoltz}
\end{figure}

In Figure~\ref{fig:validation_SymBoltz}, we display the matter power spectrum (left) and the temperature auto-correlation cosmic microwave background (right)  for several values of the coupling parameter $\gamma$ using the new \texttt{SymBoltz} code and the reference ~\texttt{CLASS} code. For the matter power spectrum, we can see that both implementations give similar results, with differences well inside the sub-percentage level for almost all scales. The relative differences remain largely independent of the value of the coupling parameter $\gamma$ indicating a code itself origin. For very large values of the coupling parameter like $\gamma \sim  10^{-6}$ differences can reach up to 2\% only for almost non-linear scales, where Boltzmann solvers are know to not be precise enough. In the case of the cosmic microwave background, we also see the results are extremely compatible, being the differences between codes below the 0.5\% level for all the modes and independent of the value of the coupling parameter. Consequently, we validated both implementation with respect to each other thanks to the similar results. They key importance of this validation is that \texttt{SymBoltz} makes use of no TCA scheme for solving the baryons-photons coupled system while \texttt{CLASS} does. In our case, we did modify the so-called \lstinline[language=bash]{TightCoupling} scheme  for \texttt{CAMB} and  the \lstinline[language=bash]{first_order_CLASS}  scheme for \texttt{CLASS} accordingly with the new terms due to the pure momentum coupling. However, we had no previous clue on whether those schemes were enough or we should move to, at least, second order. In addition to that, this interaction also provides a tight coupled system between baryons and dark matter which, in principle, could require deriving a similar TCA scheme for the dark matter-baryon system. Since \texttt{SymBoltz} is by design free from the need of those schemes, the fact that both codes gave compatible results demonstrates that the schemes used in \texttt{CLASS} and \texttt{CAMB} were enough.

\bibliography{bib}

\end{document}